\documentstyle[12pt,amstex,epsfig]{article}
\begin{document}

\begin{flushright}
MPI-PhT/97-14\\
February 1997\\
\end{flushright}
\begin{center}
\large {\bf The First Moment of $\delta g(x)$ -- a Comparative Study}
\\
\mbox{ }\\
\normalsize
\vskip1cm
{\bf Bodo Lampe}
\vskip0.2cm
Department of Physics, University of Munich\\
Theresienstrasse 37, D--80333 Munich \\
\vskip0.5cm
{\bf Andreas Ruffing}
\vskip0.2cm
Max Planck Institute for Physics \\
F\"ohringer Ring 6, D--80805 Munich \\
\vspace{1cm}

{\bf Abstract}\\                          
\end{center}                              

The sensititivity of various future polarization experiments 
to the first moment $\Delta g$ of the polarized 
gluon density is elucidated in detail. 
It is shown to what extent the first moment can be 
extracted from the future data as compared to the 
higher moments.  
We concentrate on two processes which in the near future 
will become an 
important source of information on the polarized 
gluon density, namely the photoproduction of open charm 
to be studied at CERN (COMPASS) and SLAC and the 
production of direct photons at RHIC.

\newpage                                  

{\bf 1. Introduction}

One of the main issues in polarized DIS experiments is the 
question of how the proton spin at high energies is composed 
out of the spins of its constituents, possibly 
\begin{equation}
+{1\over 2} = {1\over 2} \Delta\Sigma + \Delta g + L_z \label{111}
\end{equation}
where $\Delta\Sigma = \Delta (u+ \bar{u}) + \Delta (d+ \bar{d}) 
+ \Delta (s+ \bar{s})$ is the contribution from the quark spins. 
In the (static) constituent models, like SU(6), one has 
$\Delta\Sigma (SU_6)=1$ but experimentally 
\cite{adams} it seems that this 
rule is violated by a large amount ($\Delta\Sigma_{exp} \approx 0.25$). 
This experimental fact is also in disagreement with the 
Ellis-Jaffe sum rule \cite{ellisjaffe} which predicts 
$\Delta\Sigma_{EJ} \approx 0.65$ on the basis of the approximation 
$\Delta s = \Delta \bar{q} =\Delta g=0$. It will probably turn out 
that both the gluon and the strange quark are needed to understand 
the proton spin structure \cite{lampereya}. 

In recent studies of polarized DIS phenomena \cite{gehrmann,vogelsang} 
these questions 
have been superimposed by attempts to guess the full 
x--dependence of the polarized gluon density, but the 
question of the first moment is still particularly 
interesting because it is related to the anomaly. 
It is true that the first moment is only one among an infinite 
set of moments. However, 
the first moment $\Delta g$ certainly has its 
significance, because it enters the fundamental spin 
sum rule (\ref{111}) and because it gives the contribution 
within the proton to the $\gamma_5$ anomaly. 

In lepton--nucleon DIS the gluon 
arises only as a higher order effect 
and consequently one has
difficulties to extract the polarized gluon distribution 
and in particular its first moment 
from inclusive deep inelastic data. 
These problems have been anticipated several years ago 
by theoretical studies 
\cite{altarellistirling,carlitzreyaaltarelli,lampealtarelli}, 
and they are in fact not surprising in view of the 
subtleties in determining the unpolarized gluon density 
in unpolarized DIS experiments \cite{herastudies}. 

A popular way out of this dilemma is the study of 
semi--inclusive cross sections, and in particular of 
charm production, because the production of heavy quark hadrons  
is triggered in leading order by the photon--gluon fusion 
mechanism and is therefore sensitive to the gluon 
density inside the proton, whereas the heavy quark 
content of the proton is usually negligible at presently 
available $Q^2$--values. 
Another possibility is to look at hard prompt photons 
produced in proton--proton scattering, a process 
which is well known to lead to good results for 
$g(x)$ in unpolarized scattering.  
In this article we shall study these two processes 
in some detail and examine the 
question to what extent they are sensitive to 
the first moment $\Delta g$ as compared to the higher 
moments. 

Throughout the paper we are using polarized quark densities 
as suggested by \cite{gehrmann}. Likewise, we could 
have used densities parametrized in ref. \cite{vogelsang}. 
We stress that we make no assumptions on the form of $\delta g(x)$ 
and that our results depend only 
marginally on the quark density parametrization choosen. 

{\bf 2. Polarized Open Charm Photoproduction}

Due to its prominent decay mechanism, $J/\psi$--events 
are the most prominent within charm production, 
and this fact has led to attempts to determine the 
unpolarized gluon density from the $J/\psi$--production 
cross section \cite{bergerjones,martinryskin,halzen}. 
Similar ideas hold in the case of polarized $J/\psi$--production 
\cite{doncheski,leader3}. 
However, these suggestions are model dependent and 
depend on assumptions which go beyond the QCD improved 
parton model. For example, according to the suggestion 
of \cite{martinryskin} elastic $J/\psi$--production 
should measure the square of $\delta g(x)$   
and therefore be 
very sensitive to its magnitude.  
However, it is not clear whether in 
the cross section formula there is a factor $g(x_1)g(x_2)$ 
or whether some independent 2--gluon correlation function $K(x_1,x_2)$ 
appears.    
Furthermore, it has recently been stressed \cite{braaten}, 
that color octet contributions may appear in addition 
to the color singlet pieces in inelastic $J/\psi$--production. 
If true, this would upset the inelastic $J/\psi$--analysis 
because several new free 
parameters, the color octet matrix elements, would enter the game. 

Therefore, from the theoretical point of view 
the cleanest signal for the gluon in heavy quark 
production is probably open charm production, although 
experimentally it has worse statistics due to the  
difficulties in identifying D--mesons. 
Instead of the deep inelastic process one may as well look 
at photoproduction, because the
mass of the charm quark forces the process to take 
place in the perturbative regime. The advantage 
of photoproduction over DIS is its larger cross section. 
Two fixed target experiments, one at SLAC and COMPASS at CERN 
\cite{compass} are  
being developed to make use of this advantage 
and measure the polarized gluon 
distribution via photoproduction.  

In leading order the inclusive 
polarized {\it deep 
inelastic} open charm production cross section is given by 
\cite{reyavogelsang,watson} 
\begin{equation}
{d  \sigma_c \over dQ^2 dy}={4\pi \alpha^2 \over Q^2} 
{2-y \over yS} g_1^c({Q^2 \over yS},Q^2) 
\label{650}
\end{equation}
where S is the Mandelstam--S for the lepton--nucleon scattering 
process and  
\begin{equation}
g_1^c(x,Q^2)={\alpha_s \over 9\pi}\int_{(1+{4m_c^2 \over Q^2})x}^1 
{dw \over w} \delta g(w,Q^2)  h_{ec}({x \over w}) 
\label{651}
\end{equation}
is the charm contribution to the polarized structure 
function $g_1$ and where  
\begin{equation}
h_{ec}(z)=(2z-1)\ln {1+\beta \over 1-\beta}+(3-4z)\beta 
\label{652}
\end{equation}
is the parton level matrix element. One has  
$\beta=\sqrt{1-{4m_c^2 \over \hat s}}$ where the Mandelstam 
variable $\hat s$ is defined by $\hat s=(p+q)^2=Q^2{1-z \over z}$. 
By combining these formulae with the unpolarized cross 
section one can obtain 
the polarization asymmetries. 
If one plugs in the relatively large gluon contribution 
of references \cite{altarellistirling,gehrmann,vogelsang}, 
one gets asymmetries 
of the order 0.1 in a fixed target experiment which would 
operate well above charm threshold. 

It is straightforward to obtain from the above expressions 
(\ref{650}) -- (\ref{652}) 
the inclusive open charm {\it photoproduction} cross section 
by taking the simultaneous limits $Q^2 \rightarrow 0$ and 
$z \rightarrow 0$ while keeping ${Q^2 \over z} \approx \hat s$ 
fixed \cite{frixione,vogelsang5}: 
\begin{equation}   
\sigma^c_{\gamma p} (S_{\gamma})=
{8 \pi \alpha \alpha_s \over 9 S_{\gamma}}
\int_{{4m_c^2 \over S_{\gamma}}}^1 
{dw \over w} \delta g(w, S_{\gamma})   
(3v-\ln {1+v \over 1-v})   
\label{653}
\end{equation}     
where $v=\sqrt{1-{4m_c^2 \over \hat s}}$ and $\hat s=wS_{\gamma}$. 
This integrated cross section depends only on the 
total proton--photon energy $S_{\gamma}=(P+q)^2$ which for 
a fixed target experiment is given by $S_{\gamma}=2ME_{\gamma}$ 
where $E_{\gamma}$ is the photon energy. By varying the 
photon energy it is in principle possible to explore the 
x--dependence of $\delta g$. Very high photon energies 
correspond to small values of x. However, as we shall see 
later, it is not trivial to obtain the first moment 
of $\delta g(x)$ from the cross section Eq. (\ref{653}) 
even if the full energy dependence is known. 

It should be noted that the second argument of 
$\delta g$ in Eqs. (\ref{651}) and (\ref{653}) is not 
certain. It might as well be $4m_c^2$ or any number in 
between. This uncertainty 
reflects our ignorance about the magnitude of the 
higher order correction and could be resolved if a 
higher order calculation of these cross sections would be 
performed.  
The same statement holds true for the argument 
of $\alpha_s$. Therefore, in the equations presented below 
the energy arguments of $\delta g$ and $\alpha_s$ will be chosen 
to be more general, $\mu_S$ and $\mu_R$ respectively.  

Eq. (\ref{653}) was obtained after integration over the 
charm quark production angle ($\hat \theta$ in the gluon photon cms). 
If one is interested in the 
$p_T$ distribution or wants to introduce a $p_T$--cut, it is 
appropriate to keep the $\hat \theta$ dependence in the 
matrix element  
\begin{equation} 
ME =  
 {\hat t^2+ \hat u^2 -2m_c^2\hat s \over \hat t\hat u}
+2m_c^2{\hat t^3+\hat u^3 \over \hat t^2\hat u^2}   
\label{654}
\end{equation}
where $\hat s=wS_{\gamma}$, 
$\hat t=-{\hat s\over 2}(1-v\cos \hat \theta)$ and   
$\hat u=-{\hat s\over 2}(1+v\cos \hat \theta)$.  
It is possible to make a transformation to the transverse 
charm quark momentum by using 
$p_T^2=({\hat s \over 4}-m_c^2)\sin^2\hat \theta$ and to 
obtain the cross section for all processes with $p_T$ greater 
than a given $p_{Tcut}$ : 
\begin{eqnarray} \nonumber   
\sigma(p_{Tcut})={\pi  \alpha \alpha_s(\mu_R^2) \over 9 S_{\gamma}} 
\int_{{4(m_c^2+p_{Tcut}^2) \over S_{\gamma}}}^1 {dw \over w} 
\delta g(w,\mu_S^2) {v \over {\hat s \over 4}-m_c^2} 
\int_{p_{Tcut}^2}^{{\hat s \over 4}-m_c^2} 
{d p_T^2 \over \sqrt{1-{p_T^2 \over {\hat s \over 4}-m_c^2}}} 
\\
\biggl\{ 2m_c^2{\hat s \over \hat t\hat u}
-{\hat t \over \hat u}   -{\hat u \over \hat t}  
-2m_c^2({\hat t \over \hat u^2}+{\hat u \over \hat t^2})\biggr\}  
\label{655}
\end{eqnarray} 
There are several good reasons to study the $p_T$ distribution. 
First of all and in general, it gives more information than the 
inclusive cross section. Secondly and in particular, it can be 
shown that the integrated photoproduction cross section 
Eq. (\ref{653}) as well 
as the corresponding 
DIS charm production cross section are not sensitive 
to the first moment of $\delta g(x)$. The sensitivity is  
increased, however, if a $p_T$--cut of the order of 
$p_T \approx 1$ GeV is introduced (see below). 
Last but not least, 
it is experimentally reasonable to introduce a $p_T$--cut. 

Now we want to follow the question what the contribution of 
the first moment 
$\Delta g$ to the cross sections (\ref{651}), (\ref{653}) 
and (\ref{655}) is. 
In {\it massless} DIS this question is easy to answer. 
One can apply the convolution theorem on Eq. (\ref{651}) 
(with $m_c=0$)  
to see that the 
contribution of $\Delta g$ is given by the first moment of the 
parton matrix element. 
If masses are involved, like $m_c$, the  
answer to this question is somewhat more subtle. Since the cross 
section is not any more a convolution of the standard form,  
one can not directly apply the convolution theorem, but 
has to write it artificially as 
\begin{equation}  
\sigma (a)=\int_a^1 {dw \over w} \delta g(w) H({a\over w}) 
\label{6551}
\end{equation}  
where H is some function (to be given below) and 
$a=a_{\gamma c}={4m_c^2 \over S_{\gamma}}$ for photoproduction and 
$a=a_{ec}=(1+{4m_c^2 \over Q^2})x$ for DIS charm production.  
Now one can apply the 
convolution theorem to prove that the first moment 
$H^{(1)}=\int_0^1 dzH(z)$  
gives the contribution from $\Delta g$ to the cross section  
[and in general for the n--th moment : $\sigma^{(n)}=H^{(n)} 
\delta g^{(n)}$]. Note that the moments $\sigma^{(n)}$ are taken 
with respect to $a$.   
In DIS charm production, the function H is given by 
$H_{ec}(z) \sim  h_{ec}({z\over 1+{4m_c^2 \over Q^2}})$, 
Eq. (\ref{652}), and 
in open charm photoproduction without cuts it 
is given by $H_{\gamma c}(z)=
{8 \pi  \alpha \alpha_s(\mu_R) \over 9 S_{\gamma}}h_{\gamma c}(z)$ 
with 
\begin{equation}  
h_{\gamma c}(z)=3\sqrt{1-z} -\ln {1+\sqrt{1-z} \over 
{1-\sqrt{1-z}}} 
\label{achts}
\end{equation} 
where $z={a_{\gamma c} \over w}$.  
After some algebra one can see 
that both for the inclusive charm photoproduction 
and DIS the relevant  
quantities $\int_0^1 dzH(z)$ identically vanish. 
For example, for the photoproduction case the moment 
function $h_{\gamma c}^{(n)}$ is given by 
\begin{equation}
h^{(n)}_{\gamma c}=(n^{-1}-n^{-2})\int_0^1 (1-t^2)^n dt
\label{6541}
\end{equation}
On physical grounds the result $H^{(1)}=0$
can be traced back to the
small--$p_T$ behaviour of the matrixelement for 
$\gamma g \rightarrow c \bar c $
which cancels the contribution of the large--$p_T$ region
in $\int_0^1 dzH(z)$ \cite{schaefermankiewics}.
It is 
not really a surprise in view of the structure of the anomaly 
in massive QCD (cf. the appendix of ref. \cite{lampe4}). 

Eq. (\ref{6541}) allows to calculate $h^{(n)}_{\gamma c}$ for 
arbitrary complex n. For example, there is an expansion 
$h^{(1+\epsilon)}_{\gamma c}={2\over 3}\epsilon 
+O(\epsilon^2)$ (and similarly for $H^{(1+\epsilon)}_{ec}$) 
which shows that 
the $H^{(n)}$ keep being small 
in the neighbourhood of n=1. From this one may conclude that 
the cross sections are not suited for determining the first 
moment of $\delta g$. 

Fortunately, the situation changes 
if one includes a $p_T$--cut of greater than 1 GeV. 
In that case, some sensitivity 
to $\Delta g$ is re--established because the  
small--$p_T$ behaviour of the matrix element for 
$\gamma g \rightarrow c \bar c $ 
does not cancel the contribution of the large--$p_T$ region 
any more. 
Let us discuss this issue in some more detail for the 
photoproduction case. One can put the cross section 
(\ref{655}) in the form (\ref{6551}) with 
$H_{\gamma c}(z,p_{Tcut})=
{8 \pi  \alpha \alpha_s(\mu_R) \over 9 S_{\gamma}}
h_{\gamma c}(z,p_{Tcut})$ and  
\begin{eqnarray} \nonumber
h_{\gamma c}(z,p_{T cut})=
{v \over 8} {1 \over {\hat s \over 4}-m_c^2}
\int_{p_{Tcut}^2}^{{\hat s \over 4}-m_c^2}
{d p_T^2 \over \sqrt{1-{p_T^2 \over {\hat s \over 4}-m_c^2}}}
\biggl\{ 2m_c^2{\hat s \over \hat t\hat u}-{\hat t \over \hat u}
\\
                                    -{\hat u \over \hat t}
-2m_c^2({\hat t \over \hat u^2}+{\hat u \over \hat t^2})\biggr\}
\label{6542}
\end{eqnarray}
z is defined by $z={4m_c^2\over \hat s}$.  
All other relevant quantities have been defined before and after 
(\ref{655}).  
Note that for $p_{T cut} \rightarrow 0$  
one recovers $h_{\gamma c}(z,p_{T cut})=h_{\gamma c}(z)$, Eq. 
(\ref{achts}). 
The integral Eq. (\ref{6542}) 
will be the starting point for 
several important observations, cf. Figs. \ref{fig10}--\ref{fig215} 
below.  
The point is, that one can use the moments $H_{\gamma c}^{(n)}(p_{T cut})$ 
of $H_{\gamma c}(z,p_{T cut})$ to reconstruct the cross section  
according to the Mellin formula 
\begin{equation}
\sigma (p_{T cut})={1\over 2\pi i} 
\int_{k-i\infty}^{k+i\infty} dn a_{\gamma c}^{-n} 
H_{\gamma c}^{(n)}(p_{T cut}) \delta g^{(n)} 
\label{6544}
\end{equation}
This is because the cross section has the form Eq. (\ref{6551}) with 
$H(z)=H_{\gamma c}(z,p_{T cut})$. 
Since the integral in Eq. (\ref{6544}) extends along the imaginary 
axis, one has to consider complex values of n. 
We have studied the behavior of $H_{\gamma c}^{(n)}(p_{T cut})$ as 
a function of $n=1+iy$, $n=2+iy$, $n=3+iy$ etc.  
for real values of y and several values of $p_{T cut}$. 
In this way we are able to determine the circumstances 
under which the first moment $H_{\gamma c}^{(1)}(p_{T cut})$ 
dominates in the integral Eq. (\ref{6544}) over the 
higher moments. For definiteness, we have choosen 
a photon energy of 30 GeV ($S_{\gamma}=60$ GeV$^2$) and  
$m_c=1.3$ GeV.   

\begin{figure}
\begin{center}  
\epsfig{file=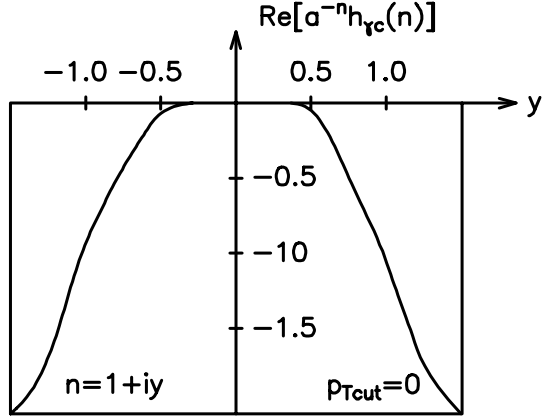,height=5.5cm,angle=0}
\bigskip
\caption{The behavior of the n--th moment of $h_{\gamma c}$ 
in the imaginary neighbourhood of $n=1$ at $p_{T cut}=0$. 
It is shown that there is no sensitivity to the first 
moment at all.}
\label{fig10}
\end{center} 
\end{figure} 

\begin{figure}
\begin{center}
\epsfig{file=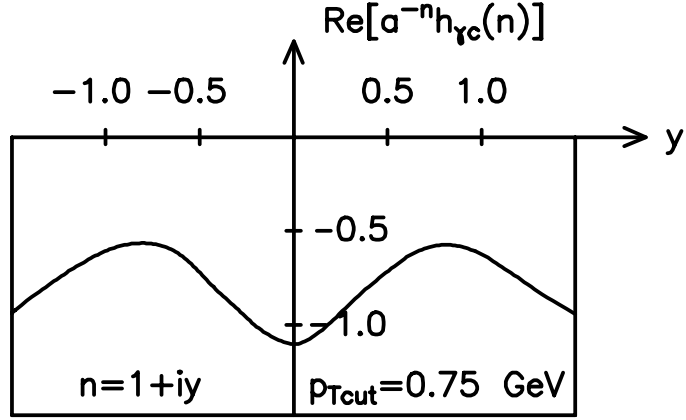,height=5.5cm,angle=0}  
\bigskip
\caption{The behavior of the n--th moment of $h_{\gamma c}$  
in the imaginary neighbourhood of $n=1$ at $p_{T cut}=0.75$
GeV. An increasing sensitivity as to the first  
moment is visible.}
\label{fig175}
\end{center}    
\end{figure} 

\begin{figure}
\begin{center}
\epsfig{file=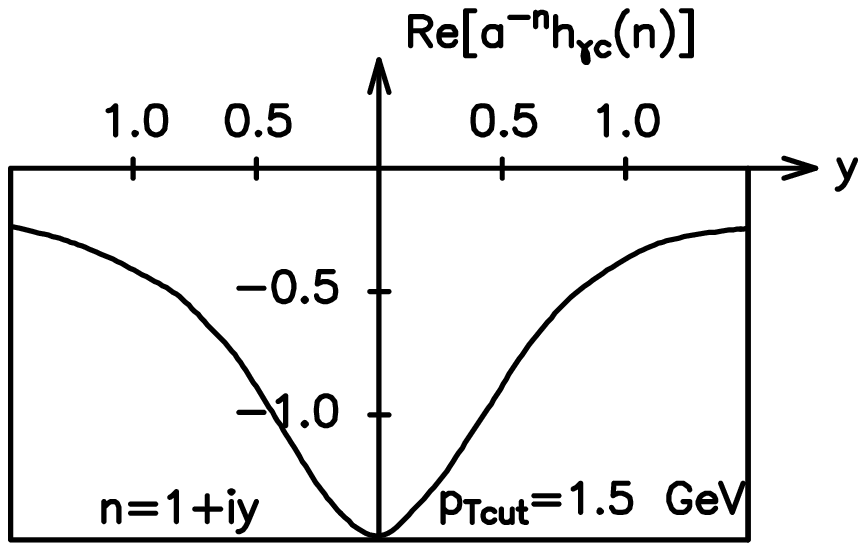,height=5.5cm,angle=0}  
\bigskip
\caption{The behavior of the n--th moment of $h_{\gamma c}$  
in the imaginary neighbourhood of $n=1$ at $p_{T cut}=1.5$   
GeV. The first moment dominates the higher moment contributions 
which are comprised at values $y \neq 0$.}
\label{fig115}
\end{center}    
\end{figure} 

More precisely, in Figs. \ref{fig10}--\ref{fig215} the quantity 
Re[$a^{-k-iy}h_{\gamma c}^{(k+iy)}(p_{T cut})$] is presented 
for $-1.5 \leq y \leq +1.5$, $k=1,2$ and $p_{T cut}=0.,
0.75,1.5$ GeV. The curves for $p_{T cut}=0$ 
(Figs. \ref{fig10} and \ref{fig20} ) can 
be obtained directly from Eq. (\ref{6541}). 
The insensitivity at $p_{T cut}=0$ as to the first 
moment is displayed by the zero of the curve 
$k=1$ (Fig. \ref{fig10}). If one compares Figs. 
\ref{fig10} and \ref{fig20}, one sees that 
the sensitivity to the second moment is much larger 
for $p_{T cut}=0$. 
The situation changes as the $p_{T cut}$ is increased. 
This can be seen in Figs. \ref{fig175}, \ref{fig115},  
\ref{fig275} and \ref{fig215}, but also in Fig. \ref{figpt} where 
$h_{\gamma c}^{(1)}(p_{T cut})$ and $h_{\gamma c}^{(2)}(p_{T cut})$ are 
given as a function of $p_{T cut}$. 
$h_{\gamma c}^{(1)}(p_{T cut})$ has an extremum at some point 
$p_{T cut} \approx 1$GeV which should be suspected to be 
the most optimal value for a measurement of $\Delta g$. 
That this is really the case, can be deducted from a 
closer study of Figs. \ref{fig175} and \ref{fig115}. 
We suggest to associate an "observational significance" 
$R_1$ to any cross section 
sensitive to $\Delta g$ by means of the following 
procedure : It should be defined as the ratio of the 
"signal" over the squareroot of the "background", 
$R_1={N_{signal}\over \sqrt{N_{background}}}$ which 
in the ideal situation of a Gaussian corresponds 
to $R_1=\sqrt{{height \over width}}$.  
In the 
case at hand one has  
$R_1 \approx {\vert H^{(1)} \vert \over 
\sqrt{\int_{\approx -1}^{\approx +1} dy a^{-1-iy}
\vert H^{(1+iy) \vert }}}$. 
Similarly, one can define a significance $R_2$ for the 
determination of the second moment and so on. 
For simplicity, $R_3$, $R_4$ etc. 
are not considered in this paper because qualitatively 
they behave not much different than $R_2$. 
From Figs.  \ref{fig10}--\ref{fig215}, 
the "observational significances" $R_1$ and $R_2$  
can be deducted as a function of $p_{T cut}$. 
If one makes a more thorough analysis by taking other 
values of $p_{T cut}$ into account, one finds that $R_1$ 
has a maximum near $p_{T cut} \approx m_c$
whereas $R_2$ goes down for increasing values of $p_{T cut}$. 
Unfortunately, $R_3$, $R_4$ etc do not decrease as strongly 
as $R_2$ at higher values of $p_{T cut}$. 
One may also determine  
the "12--significance" $R_{12}={R_1 \over R_2}$. This ratio 
is important, because it does not depend on the 
overall cross section. For example, at $p_{T cut}=0.75$ GeV one 
has $R_{12}=0.67$.  
It will be shown later that the "12--significance" 
is much larger for the RHIC process than this value obtained for 
charm photoproduction!
 
\begin{figure}
\begin{center}
\epsfig{file=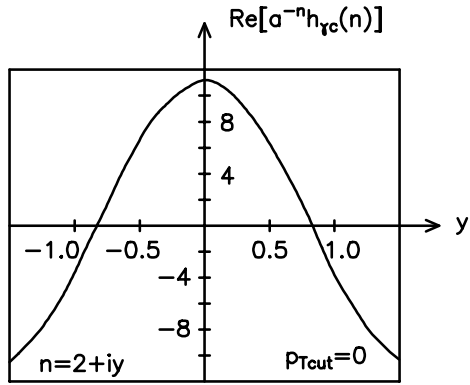,height=5cm,angle=0}
\bigskip
\caption{The behavior of the n--th moment of $h_{\gamma c}$  
in the imaginary neighbourhood of $n=2$ at $p_{T cut}=0$. 
A sensitivity of the second moment ($y=0$) as compared to the 
other moments ($y \neq 0$) is visible. The decrease at large 
values of $y>1.5$ is due to moments $n \geq 3$.} 
\label{fig20}
\end{center}
\end{figure}

\begin{figure}
\begin{center}
\epsfig{file=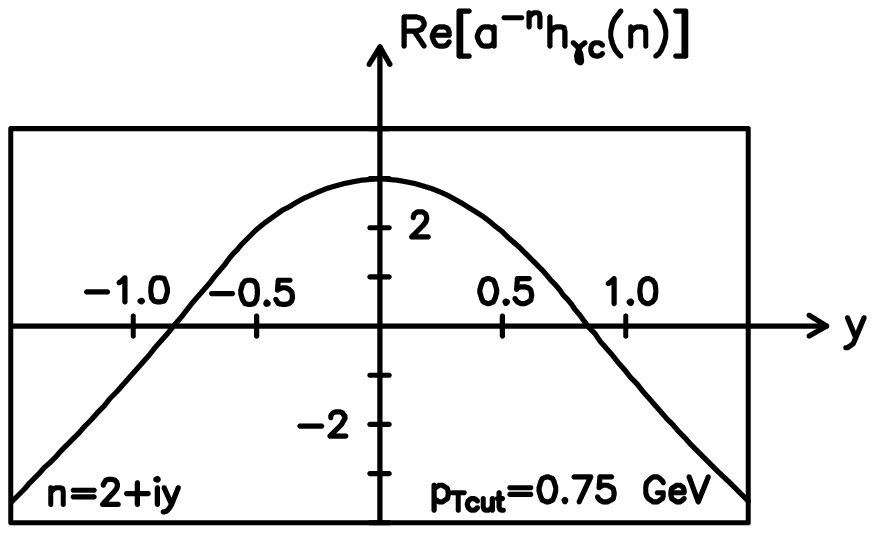,height=5cm,angle=0}
\bigskip
\caption{The behavior of the n--th moment of $h_{\gamma c}$  
in the imaginary neighbourhood of $n=2$ at $p_{T cut}=0.75$   
GeV. The sensitivity of the cross section as regards the 
second moment is going down.}
\label{fig275}
\end{center}
\end{figure}

\begin{figure}
\begin{center}
\epsfig{file=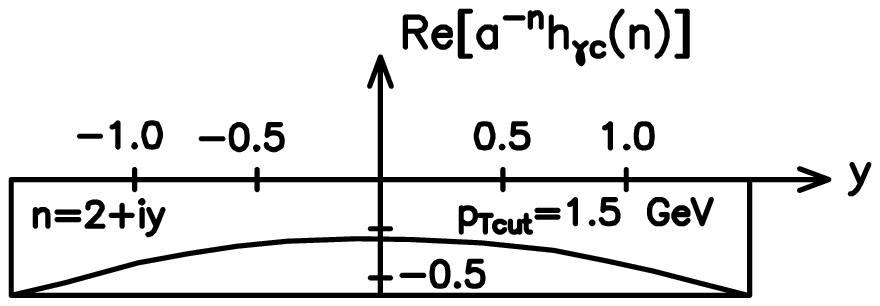,height=3.5cm,angle=0}
\bigskip
\caption{The behavior of the n--th moment of $h_{\gamma c}$  
in the imaginary neighbourhood of $n=2$ at $p_{T cut}=1.5$   
GeV. The sensitivity of the cross section as regards the  
second moment has gone down further.}
\label{fig215}
\end{center}
\end{figure}

\begin{figure}
\begin{center}
\epsfig{file=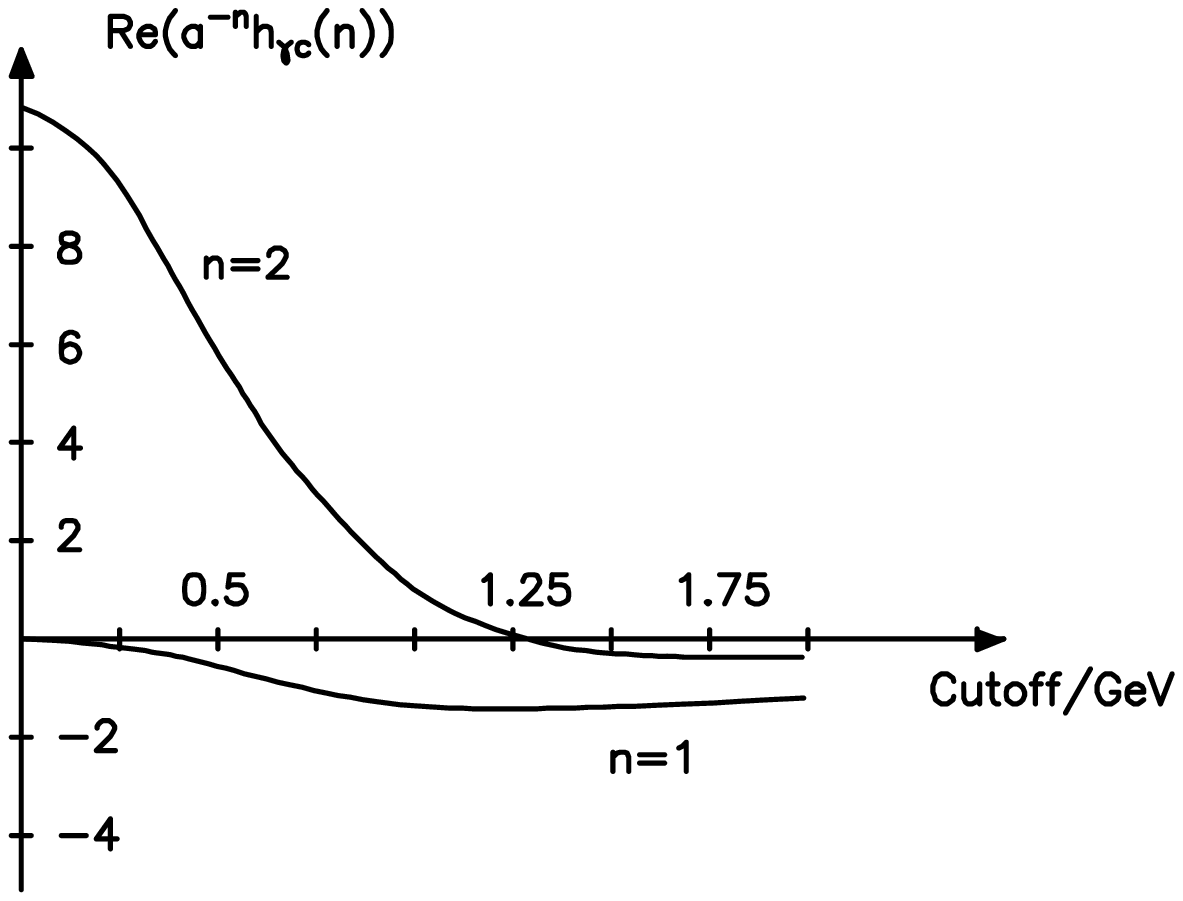,height=7.0cm,angle=0}
\bigskip
\caption{The $p_{T cut}$--dependence of  
of $h_{\gamma c}^{(n)}$ for $n=1$ and $n=2$.  
The sensitivity as regards the second moment is 
strongly reduced as the $p_{T cut}$ is switched on. 
Unfortunately, the decrease of $h_{\gamma c}^{(1)}$ 
is not as strong as one might hope.}
\label{figpt}
\end{center}
\end{figure}

{\bf 4. Polarized Production of Direct Photons in $PP$ 
          Collisions}

In unpolarized hadron scattering hard photons are known to be a clean
probe of the gluon distribution, because they can be directly detected,
without undergoing fragmentation.
Similarly, the most interesting prospect for polarized high energetic
proton experiments is the possibilitity to determine $\delta g(x)$
from the process $pp \rightarrow \gamma X$ with a high energetic
photon in the final state.
Consequently several groups have studied this process theoretically  
\cite{berger1,gupta,cheng2,bourrely1} and 
even higher order QCD corrections are completely known 
\cite{contogouris1,gordon}. 
On the experimental side, there is the upcoming very promising
experimental spin program of the Relativistic Heavy Ion Collider
(RHIC) Spin Collaboration 
\cite{yokosawa}
at the Brookhaven National
Laboratory. 
At RHIC both proton beams, with an average energy of 250 GeV each,
will be polarized, using 'Siberian snakes' , 
to an expected polarization of about 70$\%$.
Due to the high luminosity of the order of 
$\sim 10^{32}$cm$^{-2}$s$^{-2}$
(corresponding to an integrated luminosity of about 800 pb$^{-1}$)
the polarized RHIC pp collisions will play a decisive role
for measuring the polarized gluon density. 

On the parton level, direct photon production 
is induced in lowest order by
the annihilation of light quarks $q \bar{q} \rightarrow \gamma g$ and
by the Compton scattering
$qg \rightarrow \gamma q$. Among the two, 
the contribution from the annihilation process is small 
because it is a valence--sea scattering process. 
Since the Compton scattering dominates,  
the cross section is strongly dependent on the
magnitude of  $\delta g(x)$. Assuming the above luminosity at RHIC,  
a sensitivity of about 5$\%$
is expected for ${\delta g(x) \over g(x)} $ \cite{rhicdg}. 
If $\delta g(x)$ is large, one encounters
large positive values of the spin asymmetry up to 50 percent. 
The form of $\delta g(x)$ can largely be reconstructed from the 
$k_T$--distribution of the direct photons which is given by the 
simple formula
\begin{equation}
{d \sigma (PP \rightarrow \gamma X) \over da}=
\sum_{qg,gq}
\int_a^1 dw  \delta g(w) \int_{a \over w}^1 dx  
{d \hat{\sigma} (qg/gq \rightarrow \gamma q)\over da} 
\sum_q Q_q^2 \delta q(x)
\label{671}
\end{equation}
where $a=a_{pp}={4 k_T^2 \over S}$ is the rescaled transverse momentum 
and 
the parton level $k_T$--distribution 
${d  \hat{\sigma} \over da}
={1 \over 2\vert c\vert x w } {d  \hat{\sigma} \over d\vert c\vert }$ 
is given by 
\begin{equation}
{d \hat{\sigma} (qg(gq) \rightarrow \gamma q) \over dc}=
{\pi \alpha \alpha_s \over 6 xwS}
[ {1 \over {1 \over 2} (1 \pm c)} - 
{1 \over 2} (1 \pm c)]
\label{672}
\end{equation}
The upper and lower sign stand for the processes 
$qg \rightarrow \gamma q$ and $gq \rightarrow \gamma q$, 
respectively. 
c is the cosine of the parton--parton scattering 
angle in the parton cms. Note that the photon transverse 
momentum $k_T^2={1\over 4}\hat s (1-c^2)={1\over 4}xwS (1-c^2)$ 
is identical for parton and proton 
level.

\begin{figure}
\begin{center}
\epsfig{file=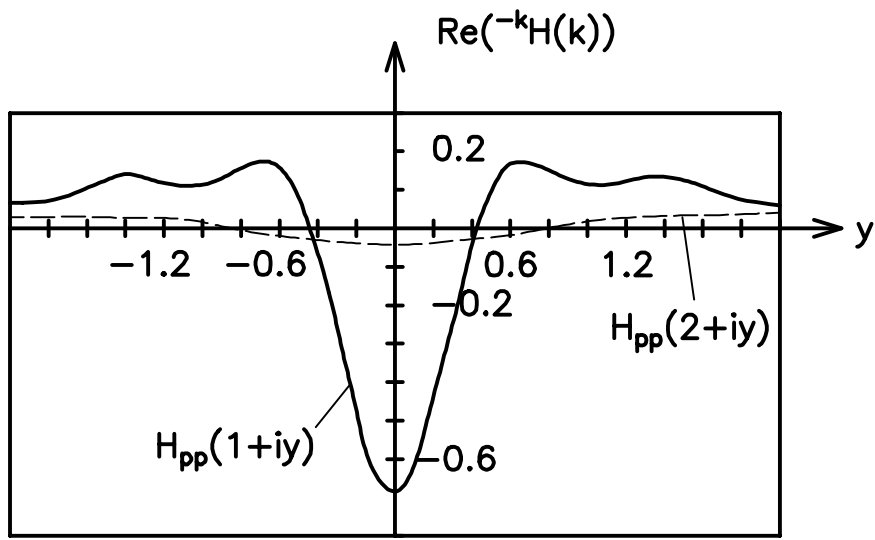,height=7.0cm,angle=0}
\bigskip
\caption{The behavior of the n--th moment of $H_{pp}$
in the imaginary neighbourhood of $n=1$ and $n=2$. 
The dominance of the first moment over the higher 
moment is clearly visible. It is much more pronounced 
than in any of the photoproduction cases considered before. 
Numbers for $H_{pp}$ are given in units of 
${\pi \alpha \alpha_s \over 6 a_{pp}   S}$. }
\label{figpp}
\end{center}
\end{figure}

The question then arises, how sensitive this 
measurement is to the first moment of  $\delta g(x)$ as 
compared to the higher moments. This question can 
be answered in a similar fashion than for charm 
photoproduction. The main ingredient is again the Mellin 
theorem which can be applied to the photon $k_T$--distribution 
Eq. (\ref{671}) because 
${d \sigma (PP \rightarrow \gamma X) \over da_{pp}}$
is of the form 
$\int_{a_{pp}}^1 {dw \over w} \delta g(w) H_{pp}({a_{pp}\over w})$
where $H_{pp}(z)=H_{pp}^+(z)+H_{pp}^-(z)$ is the sum of 
contributions from the processes 
$qg \rightarrow \gamma q$ and $gq \rightarrow \gamma q$. 
\begin{equation}
H_{pp}^{\pm}(z)={ \pi \alpha \alpha_s \over 12 a_{pp}  S}
\int_z^1 {dx \over x} [Q_u^2 \delta u(x)+Q_d^2 \delta d(x)] 
 {z\over x \vert c\vert } 
[ {1 \over {1 \over 2} (1 \pm c)} - 
{1 \over 2} (1 \pm c)]
\label{673}
\end{equation}
where $c = \pm \sqrt{1-{z\over x}}$. 
Therefore, 
according to the convolution theorem, the contribution 
of $\Delta g$ to the cross section is given by the 
first moment of $H_{pp}(z)$. 
Introducing $u={z \over x}$, it can be seen that the 
moments of $H_{pp}(z)$ factorize in a very simple way 
\begin{equation}
H_{pp}^{(n)}={\pi \alpha \alpha_s \over 6 a_{pp} S}
\int_0^1 du u^{n-1} ({u\over 2}+2) 
\int_0^1 dx x^{n-1} [Q_u^2 \delta u(x)+Q_d^2 \delta d(x)]
\label{674}
\end{equation}
Just as in the case of charm photoproduction we can now 
study the behavior of $H_{pp}^{(n)}$ in the imaginary neighbourhood 
of $n=1$ and $n=2$, and in principle also for higher moments. 
The results are shown in Fig. \ref{figpp} and exhibit a 
distinct peak at $n=1$ whereas the sensitivity to the 
second moment is much weaker (almost flat). The peak is in fact much 
more pronounced than any of the peaks which were found 
in the last section for charm photoproduction. 
From the figure we calculate 
the 12--sensitivity defined in the last section to be 
$R_{12}={0.175 \over 0.042}=4.2$. Similar large  
values arise for $R_{13}={R_1 \over R_3}$, 
$R_{14}={R_1 \over R_4}$ etc. Therefore one 
concludes that this process is much more sensitive 
to  $\Delta g$ than polarized charm photoproduction. 

Higher order QCD corrections to the process                
$pp \rightarrow \gamma X$ involving polarized proton beams 
have been calculated in refs. \cite{contogouris1,gordon}.  
It is possible to extend our method to include the higher 
order effects because the results of the calculation 
are given in the form of a K--factor to the 
$k_T$ distribution. 
The higher order corrections turn out to be  
positive          
and quite large. 
Nevertheless, the "12--significance" is hardly 
modified because the higher order effects cancel between 
the numerator $R_1$ and denominator $R_2$ (cf. the end of section 3).               

There is another RHIC process which will serve to give 
information about $\delta g(x)$, namely the production of 
heavy quarks in the collision of polarized protons
\cite{yokosawa}.  Using our method it is possible 
to determine its sensitivity to $\Delta g$. At low and 
intermediate values of the heavy quark $p_T$, the 
cross section is dominated by the subprocess $gg \rightarrow Q \bar Q$. 
Correspondingly, the moments of the cross section are generically 
of the form $\sigma^{(n)} =[\delta g^{(n)}]^2 \hat \sigma^{(n)}$. 
Due to the quadratic dependence on $\delta g^{(n)}$ one concludes 
that this process is very sensitive to $\Delta g$ if 
$\Delta g$ is large and not very sensitive to $\Delta g$ 
if $\Delta g$ is small. The 'weight' $\sigma^{(1)}$ can be  
calculated to be relatively large. However, a quantitative 
comparison to the results obtained above is not possible, 
because direct photon production is linear in $\delta g(x)$ 
and heavy quark production is quadratic in $\delta g(x)$ and 
because the magnitude of $\Delta g$ is not known.

{\bf 5. Summary}

Present experimental data do not really give good information 
about the polarized gluon density $\delta g(x)$, 
\footnote{
It has been attempted to
determine the magnitude of the first moment $\Delta g$ \cite{forte}
from the DIS data. As an essential
input the presently known $Q^2$--dependence of the data (both theoretically
and experimentally) was used, in which the polarized gluon plays some
role. On the basis of this, a value $\Delta g \approx 1.3 \pm 0.5$ was
quoted. Our opinion is that the error here is underestimated but the
order of magnitude of $\Delta g$ looks reasonable. 
Note there are also attempts to determine 
$\Delta g$ from meson properties, e.g. recently ref. \cite{fritzsch}. }
because in DIS the gluon is a higher order effect. 
However, in the near future several experiments at BNL, CERN and 
SLAC will directly test for $\delta g(x)$. In this paper 
we have examined the question of how sensitive these 
experiments will be to $\Delta g$ as compared to the 
higher moments. We have found that for the determination 
of $\Delta g$ the RHIC experiment 
is suited much better than the charm photoproduction 
process. This arises due to the property of the matrix elements 
and is not just because RHIC allows to study 
smaller values of x, but because the full cross section 
gets a relatively larger contribution from the first 
(as compared to the higher) moments.    
Using our method it is in principle possible to 
separately analyze the sensitivity to 
any moment, but we have restricted ourselves to the 
first  
moment because of its  
outstanding importance. 
For comparative reasons we have included in our analysis 
the second moment as well. 

In principle, to determine the first moment 
$\Delta g$ precisely it is necessary to 
know the small--x behaviour $\delta g(x)$. 
In reality, all processes allow to determine 
$\delta g(x)$ only down to some lower limit $x \geq a$. 
For example, in charm photoproduction one has 
$x \geq {4m_c^2 \over S_{\gamma}}$ 
for kinematical reasons. Thus, increasing the photon energy 
from 30 ($x \geq 0.12$) to 300 GeV ($x \geq 0.012$), 
one can penetrate deeper into the 
small--x region, and so on. 
Clearly, working at a fixed photon energy (fixed a), 
one will not get any information on $\delta g(x \leq a)$. 
As a consequence, the moments determined from $\delta g(x)$ 
will be "cut moments". For example, the first cut moment 
will be $\Delta_a g :=\int_a^1 dx\delta g(x)$. 
We want to make clear that the whole analysis presented 
in this article refers to the first cut moment. 
One has $a={4m_c^2 \over S_{\gamma}}$ and   
$a=4{m_c^2+p_{Tcut}^2 \over S_{\gamma}}$ for 
charm photoproduction with and without $p_{Tcut}$,  
respectively. Thus,  
application of the $p_{T}$--cut effectively increases $a$ 
and shrinks the x--region which can be studied.  
However, 
we have seen in Sect. 3 that one has to apply a $p_{T}$--cut 
because otherwise there would be no 
sensitivity to the first moment at all. 
On the other hand, our results apply to cut moments 
only up to terms of order $O(a)$ because the convolution 
formula is true strictly speaking for moments and not 
for cut moments. In charm photoproduction we studied 
the case $a=0.12$ (cf. Figs. \ref{fig10}--\ref{fig215}), 
so that we expect corrections 
of about 10$\%$ to our results. 

The situation is much better for the production of direct 
photons at RHIC. First of all, as shown in Sect. 4, 
there is a stronger sensitivity to the first moment 
as in charm photoproduction, 
and secondly, it will be possible to penetrate deeper  
into the small--x region, because in this case 
$a={4k_T^2 \over S}$. It is true that hard photons 
with transverse energies less than a few GeV are difficult to 
identify, but even if one considers only photons with 
$k_T \geq 10$ GeV, one still has $a \geq 0.0016$, and 
the first cut moment $\Delta_{0.0016} g$ will contain 
a lot of contributions from the small--x regime.  

Recently, the small--x behavior of the polarized parton 
densities have attracted some attention. 
We think that it can safely be 
stated that the polarized gluon density behaves much more 
moderate than the unpolarized one. It can be shown on rather 
general grounds that $\delta g(x)$ is one power of x less 
singular than $g(x)$. Some results in the literature \cite{ryskin}
indicate that logarithms of x may be present and 
consequently   
predict a relatively strong rise of $\delta g(x)$ 
as $x \rightarrow 0$. These results are not 
really indicative for the first moment, 
because they neglect certain contributions at very small 
x which are assumed to 
compensate those logarithms \cite{bartelsprivatecommunication}. 
In fact, the first moment of $g_1$ would not exist if 
one would take the results of 
\cite{ryskin} literally.  
They are, however,  a good starting point for 
measurements of $g_1(x)$ in the small--x regime 
feasibale at HERA. Maybe, they will already be seen at RHIC.

\end{document}